\documentclass[a4paper,10pt]{article}
\usepackage{graphicx}
\usepackage{amsmath}
\usepackage{amsfonts}
\usepackage{amssymb}
\usepackage{epsfig}
\newcommand{\be}{\begin{equation}}
\newcommand{\ee}{\end{equation}}

\newcommand{\no}{\nonumber\\}
\newcommand{\ba}{\begin{eqnarray}}
\newcommand{\ea}{\end{eqnarray}}
\newcommand{\bg}{\begin{multline}}
\newcommand{\eg}{\end{multline}}
\def\gl#1{(\ref{#1})}
\def\tr#1{\mbox{\rm tr}\left\{#1\right\}}

\newcommand{\la}[1]{\label{#1}}

\renewcommand{\thefootnote}{\fnsymbol{footnote}}
\begin{document}

\centerline{\Large\bf On the possibility of $P$-violation  at finite baryon-number densities}
\bigskip

\centerline{\large A.A.~Andrianov\footnote{ On leave of absence from V.A. Fock Department of Theoretical Physics, St. Petersburg State University,
 Russia;\\ andrianov@ecm.ub.es} and D.~Espriu\footnote{espriu@ecm.ub.es}}
\medskip

\centerline{\it Departament d'Estructura i Constituents de la Mat\`eria and
 Institut de Ci\`encies del Cosmos,}\centerline{\it Universitat de Barcelona,
 Diagonal 647, 08028 Barcelona, Spain}

\setcounter{footnote}{0}
\renewcommand{\thefootnote}{\arabic{footnote}}

\begin{abstract}
We show how the introduction of a finite baryon density may trigger spontaneous parity violation in the hadronic phase of QCD. Since this involves strong interaction physics in an intermediate energy range we  approximate QCD by a $\sigma$ model that retains the two lowest  scalar and pseudoscalar multiplets. We propose a novel mechanism based on interplay between lightest and heavy meson states which cannot be realized solely in the Goldstone boson (pion) sector and thereby is unrelated to the one advocated by Migdal some time ago.  Our approach is relevant for dense matter in an intermediate regime of few nuclear densities where quark percolation does not yet play a significant role.
\end{abstract}


\section{\label{intro} Introduction}
Some time ago it was proved quite rigorously in \cite{witten} that parity, $P$, and vector flavor symmetry could not undergo spontaneous symmetry breaking in a vector like theory such as QCD. This is thus a well established result in strong interactions at {\it zero} chemical potential. Finite baryon density however results in a manifest breaking of $CP$ invariance. The presence of a finite chemical potential leads to the presence of a {\it constant} imaginary zeroth-component of a vector field and the partition function of QCD is not anymore invariant under a $CP$ transformation. The conditions under which the results of \cite{witten} were proven (positivity of the measure) then do not hold anymore.

The appearance of $P$ violation for sufficiently large values of the chemical potential, i.e. at finite baryon density it is thus a logical possibility (conjectured by \cite{migdal} in nuclear physics long ago). Can this possibility be realized in nature? Which would be its observable consequences should it occur?

In order to answer the first question we could always appeal to lattice QCD for help. In fact, this possibility has been studied intensively for quite some time. However, finite density simulations are notoriously difficult (the fermion determinant is in general complex for a non-zero chemical potential $\mu$) and one has to resort to more involved techniques such as determining the phase of the determinant separately,  Taylor expansions in $\mu$ or analytic continuation to imaginary chemical potential. On top of that Pauli blocking is at work for large values of $\mu$, etc. For a comprehensive review of the ongoing issues, see for instance \cite{philip}. Thus the lattice results for sufficiently large values of the baryon chemical potential (where the effect is expected to appear) are not known rigorously yet\footnote{The existence of a lattice parity breaking phase at strong coupling --the so-called Aoki phase- is well established and it is due to the fact that even at vanishing chemical potential the discretization of the fermion action leads to non-positive determinants\cite{aoki}. This is clearly a lattice artifact and it is not the effect we are after.}.

It is simpler to consider the (less relevant for physics) case of  isospin chemical potential at zero baryon density \cite{son} . In this case the fermion determinant is positive and spontaneous parity breaking is excluded by the Vafa-Witten theorem, although a pion condensate is in principle still possible.

In this work we shall attempt to explore the interesting issue of $P$-parity  breaking employing effective lagrangian techniques in the range of nuclear densities for which hadron phase persists and quark percolation does not occur yet\footnote{A qualitative estimate of the densities which enforce the overlapping of pions clouds, but not the cores, of nucleons indicates the necessity of including heavy meson states.}. Our effective lagrangian is a generalized linear $\sigma$ model where we only include the lowest lying resonances, those that are expected to play a role in this issue.  The use of effective Lagrangians is also crucial to answer the second question of interest, namely how would parity breaking originating from a finite baryon density eventually reflect in hadronic physics.

Let us mention here several possible signatures of $P$-parity breaking. \\
a) Decays of higher-mass meson resonances (radial excitations) into pions. Resonances do not have a definite parity and therefore the same resonance can decay both in two and three pions (in general into even and odd number of pions).\\
b) At the very point of the phase transition leading to parity breaking one has {\it six} massless pion-like states. After crossing the phase transition, in the parity broken phase, the massless charged pseudoscalar states remain as Goldstone bosons enhancing  charged pion production, whereas the additional neutral pseudoscalar state becomes massive.\\
c) Reinforcement of long-range correlations in the pseudoscalar channel and, correspondingly,
additional isospin breaking effects in the pion decay constant and substantial modification of $F_{\pi'}$ for massless charged pions, giving an enhancement of electroweak decays.

In the next section we shall introduce the model and see that there are some subtleties associated to the choice of the low-energy effective hadronic theory. Too simple models are not rich enough to explore all the different phases that the presence of manifest $CP$ violation opens for us. We shall impose on the model the conditions for it not to lead to spontaneous breaking of parity for vanishing baryon chemical potential. In section \ref{finite-mu} we shall introduce the finite chemical potential and see how it modifies the effective theory and the vacuum state. Next, in section \ref{spectrum} we shall understand how the masses and couplings of the particles are modified in such conditions. A specific model, that describes low-energy QCD,  is studied in section \ref{Z-symmetry} and it is seen that the values of the low-energy constants in QCD are compatible with the emergence of the phenomenon of spontaneous symmetry breaking at intermediate densities ranging from 3 to 5 times the normal nuclear density, approximately.
The range of intermediate nuclear densities is of high interest as they may be reached in both   compact stars \cite{itoh,grei} and  heavy-ion collisions \cite{heavyion} . 

There are some previous studies dealing with the problem of strong interactions at zero temperature and finite chemical potential: depending on a value of nuclear density, a variety of methods  are involved using meson-nucleon  \cite{migdal,bary} or quark-meson \cite{grei,sigqu} Lagrangians and  models of Nambu-Jona-Lasinio type \cite{NJL,scon}. One has also to mention an adjacent phenomenon of $(C)P$-parity  breaking in meta-stable nuclear bubbles created in
hot nuclear matter \cite{kharzeev}.

\section{\label{sigma-model} Generalized sigma model}

The simplest hadronic effective theory is the  linear $\sigma$-model of Gell-Mann and Levy \cite{gellmann}, which contains a multiplet of the lightest scalar $\sigma$ and pseudoscalar $\pi^a$ fields. Spontaneous chiral symmetry breaking  emerges due to a non-zero value for $\langle \sigma\rangle  \sim \langle \bar q q \rangle /
 F^2_\pi$ . Current algebra techniques indicate that in order to relate
this model to QCD one has to choose a real condensate for the scalar density, with its sign opposite to current quark masses, and avoid any parity breaking due to a v.e.v. of the pseudoscalar density. The introduction of a chemical potential does not change the phase of the condensate and therefore does not generate any parity breaking. This is just fine because in normal conditions $P$ breaking is impossible in QCD.

However, if two different scalar fields condense with a relative phase between the two v.e.v.'s the opportunity of spontaneous parity breaking may arise.

Let us consider a model with two multiplets of scalar/pseudoscalar fields \be H_j = \sigma_j {\bf I} + i \hat\pi_j, \quad j = 1,2;\quad H_j H_j^\dagger = (\sigma^2_j + (\pi^a_j)^2 ) {\bf I} , \ee with $\hat\pi_j \equiv \pi^a_j \tau^a$ with $\tau^a$ being a set of Pauli matrices. We shall deal with a scalar system globally symmetric in respect to $SU(2)_L \times SU(2)_R$ rotations and work in the exact chiral limit, with vanishing current quark masses. We should think of these two chiral multiplets as representing the two lowest-lying radial states\footnote{Some previous attempts to give the Lagrangian description for two multiplets of scalar and pseudoscalar mesons have been undertaken in \cite{bars} (the oldest one ) and in \cite{schech} (the most recent one). However we have been rather inspired by our previous works on extended quark models \cite{ava,aet}. } for a given $J^{PC}$ . Of course one could add more multiplets, representing higher radial and spin excitations, to obtain a better description of QCD, but the present model already possesses all the necessary ingredients to study spontaneous parity breaking (SPB).  Inclusion of higher-mass states is required at substantially larger densities when typical distances between baryons are shrinking considerably and meson excitations  with Compton wave lengths much shorter than a pion one  play an important role.

Let us define the effective potential of this generalized $\sigma$ model. First we write the most general Hermitian potential at zero $\mu$ compartible with $SU(2)_L \times SU(2)_R$
\ba V_{\text{eff}}&=& \frac12 \tr{- \sum_{j,k=1}^2 H^\dagger_j \Delta_{jk} H_k + \lambda_1 (H^\dagger_1 H_1)^2 + \lambda_2 (H^\dagger_2 H_2)^2+ \lambda_3 H^\dagger_1 H_1 H^\dagger_2 H_2 \right.\no && \left.+ \frac12 \lambda_4 (H^\dagger_1 H_2 H^\dagger_1 H_2 + H^\dagger_2 H_1 H^\dagger_2 H_1) + \frac12 \lambda_5 (H^\dagger_1 H_2 + H^\dagger_2 H_1) H^\dagger_1 H_1  \right.\no &&\left.+ \frac12 \lambda_6 (H^\dagger_1 H_2 + H^\dagger_2 H_1) H^\dagger_2 H_2  } + {\cal O}(\frac{|H|^6}{\Lambda^2}), \la{effpot1} \ea with 9 real constants $\Delta_{jk}, \lambda_A$ . QCD bosonization rules indicate that $\Delta_{jk}\sim \lambda_A \sim N_c$. The neglected terms will be suppressed by inverse power of the chiral symmetry breaking (CSB) scale $\Lambda \simeq 1.2$GeV. If we assume the v.e.v. of $H_j$ to be of the order of the constituent mass $0.2 \div 0.3$ GeV,  it is reasonable to neglect these terms. In section \ref{Z-symmetry} we shall estimate the values of the constants $\lambda_A$ approximately corresponding to QCD.

After specifying the v.e.v. $\langle H_1\rangle = \langle\sigma_1\rangle$, one can use the global invariance of the model to factor out the Goldstone boson fields
 with the chiral parameterization
\ba H_1 (x) = \sigma_1 (x) U(x) = \sigma_1 (x) \xi^2(x);\quad H_2 (x) = \xi (x)\big(\sigma_2 (x) +i \hat\pi_2 (x)\big)\xi (x). \label{chipar}\ea
This kind of parameterization preserves the parities of $\sigma_2 (x)$ and $\hat\pi_2$ to be even and odd respectively (in the absence of SPB) and realizes manifestly the masslessness of Goldstone bosons.

Let us now investigate the hypothetical appearance of a non-zero v.e.v. of pseudoscalar fields. In order not to break the charge conservation, we must expect, if at all,  only a neutral condensate represented by a solution with $\pi_2^a= \delta^{a0} \rho$. The conditions to have an extremum are derived from the first variation of the effective potential \gl{effpot1} after substitution of Eq. \gl{chipar}, \ba &&2(\Delta_{11} \sigma_1 + \Delta_{12} \sigma_2) = 4 \lambda_1 \sigma_1^3 + 3\lambda_5 \sigma_1^2 \sigma_2 + 2 (\lambda_3 + \lambda_4) \sigma_1 \sigma_2^2  + \lambda_6  \sigma_2^3 + \rho^2 \Big(2(\lambda_3 - \lambda_4) \sigma_1 + \lambda_6  \sigma_2\Big) , \no &&2(\Delta_{12} \sigma_1 +\Delta_{22} \sigma_2) = \lambda_5 \sigma_1^3 + 2 (\lambda_3 + \lambda_4)  \sigma_1^2 \sigma_2 +   3 \lambda_6  \sigma_1 \sigma_2^2 + 4 \lambda_2 \sigma_2^3  + \rho^2 \Big(\lambda_6  \sigma_1 +4 \lambda_2 \sigma_2 \Big), \no &&0 = 2\pi^a_2\Big( - \Delta_{22} +(\lambda_3 - \lambda_4) \sigma_1^2 + \lambda_6 \sigma_1 \sigma_2+ 2\lambda_2 \sigma_2^2 + 2  \lambda_2 \rho^2\Big) .\la{efeqs} \ea
To avoid spontaneous parity breaking in normal vacuum phase of QCD, it is {\it necessary and sufficient} to impose \be (\lambda_3 - \lambda_4) \sigma_1^2 + \lambda_6 \sigma_1 \sigma_2 + 2\lambda_2 \sigma_2^2  > \Delta_{22} . \la{ineq1} \ee Since QCD in normal conditions does not lead to parity breaking, the low-energy model must necessarily fulfill \gl{ineq1}.

A necessary condition to have a minimum for non-zero $\sigma_j$ (for vanishing $\rho$) can be
 derived from the condition to get a local maximum (or at least a saddle point)
 for zero $\sigma_j$.  The sufficient conditions follow from the positivity of the second variation for a non-trivial solution of the two first equations \gl{efeqs} at $\rho = 0$.

\section{\label{finite-mu}Finite chemical potential}

We shall assume that the scalars under consideration are  generated in the quark sector of QCD. The baryon chemical potential is transmitted to the meson sector via a quark-meson coupling.  Without loss of generality we can choose the collective field having local coupling to quarks as $H_1$; this actually defines the chiral multiplet $H_1$. The set of coupling constants in \eqref{effpot1} is sufficient to support this choice as well as to fix the Yukawa coupling constant to unity. Thus finite density is transmitted to the boson sector via \be \Delta {\cal L}= - (\bar q_R H_1 q_L +\bar q_L H_1^\dagger q_R), \ee where $q_{L,R}$ are assumed to be constituent quarks. Then the one-loop contribution to $V_{\text{eff}}$ is \ba \Delta V_{\text{eff}}(\mu) &=& \frac{{\cal N}}{2} \Theta(\mu-|H_1|)\left[\mu|H_1|^2\sqrt{\mu^2-|H_1|^2}- \frac{2\mu}{3}(\mu^2-|H_1|^2)^{3/2}- |H_1|^4\ln{\frac{\mu+\sqrt{\mu^2-|H_1|^2}}{|H_1|}}\right] \no &&\times\Big(1+O\left(\frac{\mu^2}{\Lambda^{2}}; \frac{|H_1|^2}{\Lambda^{2}}\right)\Big);\qquad
 {\cal N} \equiv \frac{N_{c}N_f}{4\pi^2},\label{potmu} \ea where $\mu$ is the chemical potential.
 The higher-order contributions of chiral expansion in $ 1/\Lambda^2$  are not considered.
 This effective potential is normalized to reproduce the baryon density
for quark matter \ba \rho_B = -\frac13 \partial_\mu \Delta V_{\text{eff}}(\mu) = \frac{N_{c}N_f}{9\pi^2} p_F^3 = \frac{N_{c}N_f}{9\pi^2} (\mu^2-|\langle H_1\rangle|^2)^{3/2}, \ea where the quark Fermi momentum is $p_F = \sqrt{\mu^2-|\langle H_1\rangle |^2}$. Normal nuclear density is $\rho_B \simeq 0.17$ fm$^{-3} \simeq (1.8$ fm$)^{-3}$ that corresponds to the average distance $1.8$ fm between nucleons in nuclear matter.

After the introduction of $\mu$ the conditions for a minimum of the effective potential are modified. In particular, the first equation in \gl{efeqs} takes the form  \ba 2(\Delta_{11} \sigma_1 + \Delta_{12} \sigma_2) &=& 4 \lambda_1 \sigma_1^3 + 3\lambda_5 \sigma_1^2 \sigma_2 + 2 (\lambda_3 + \lambda_4) \sigma_1 \sigma_2^2  + \lambda_6  \sigma_2^3  + \rho^2 \Big(2(\lambda_3 - \lambda_4) \sigma_1 + \lambda_6  \sigma_2\Big)\no && + 2 {\cal N}\Theta(\mu-\sigma_1)\left[\mu \sigma_1\sqrt{\mu^2-\sigma_1^2}- \sigma_1^3\ln{\frac{\mu+\sqrt{\mu^2-\sigma_1^2}}{\sigma_1}}\right], \label{efeqs1}\ea
and other ones are unchanged.

Let us first focus on the regime of small chemical potentials. For small values of $\mu$, we know that the value of the odd parity condensate $\rho$  is zero. 
The possibility of SPB is controlled by the inequality \gl{ineq1}; in order to approach a SPB phase transition we have to diminish the l.h.s. of inequality \gl{ineq1} and therefore we need to have (assuming that the inequality indeed holds at $\mu=0$) \be
\partial_\mu\Big[(\lambda_3 - \lambda_4) \sigma_1^2
+ \lambda_6 \sigma_1 \sigma_2 + 2\lambda_2 \sigma_2^2 \Big] < 0. \la{ineq3}\ee

Let us now leave the case $\mu\simeq \langle\sigma_1\rangle$ and examine the possible existence of a critical point where the strict inequality \gl{ineq1} does not hold and instead for $\mu > \mu_{crit}$ \be (\lambda_3 - \lambda_4) \sigma_1^2 +  \lambda_6 \sigma_1 \sigma_2 +2\lambda_2 \Big(\sigma_2^2    + \rho^2\Big) = \Delta_{22} . \la{creq1} \ee  After substituting $\Delta_{22}$ from \gl{creq1} into the second equation in \gl{efeqs} one finds that \be
  \lambda_5 \sigma_1^2 + 4 \lambda_4 \sigma_1 \sigma_2+
\lambda_6 \Big(\sigma_2^2    + \rho^2\Big)  = 2\Delta_{12},\label{creq2} \ee where we have taken into account that $\sigma_1\neq 0$. Together with \gl{creq1} this completely fixes the v.e.v.'s of the scalar fields $ \sigma_{1,2}$. If $\lambda_2 = 0$ and/or $ \lambda_6=0$ the corresponding Eq. \gl{creq1} or \gl{creq2} firmly fix
 the relation between  and $\sigma_2$.
Otherwise if  $\lambda_2 \lambda_6 \not=0$ these two equations still allow to get rid of the v.e.v. of pseudoscalar field.  Thus in the $P$-breaking phase the relation between the two scalar v.e.v's is completely determined.
Using Eqs. \gl{efeqs}, \gl{efeqs1} and \gl{creq2}  one can easily eliminate the variables $\rho$ and $\sigma_2$ and get for numerical calculations the equation for $\sigma_1$ solely.

Let us now try to determine the critical value of the chemical potential, namely the value
 where $\rho (\mu_c) = 0$, but Eqs.\gl{creq1},
\gl{creq2} hold. Combining these two equations, \ba (4\lambda_2 \Delta_{12} - \lambda_6 \Delta_{22}) r^2 + (2\lambda_6 \Delta_{12} - 4\lambda_4 \Delta_{22}) r + 2(\lambda_3 - \lambda_4)\Delta_{12}- \lambda_5 \Delta_{22} = 0;\quad r \equiv \frac{\sigma_2}{\sigma_1}. \label{homeq} \ea In order for a SPB phase to exist this equation has to possess real solutions. If $4\lambda_2 \Delta_{12} - \lambda_6 \Delta_{22}\neq 0$ the SPB phase is bounded by two critical points corresponding to second order transitions. If, on the contrary, $4\lambda_2 \Delta_{12} - \lambda_6 \Delta_{22}= 0$ there is only one solution corresponding to a second order transition, but there may exist other solutions that fall beyond the accuracy of our low energy model (which becomes inappropriate for small values of $\sigma_1$). See section \ref{Z-symmetry}.

\section{\label{spectrum} The physical spectrum in the SPB phase}

Once a condensate for $\pi^{0}_2$ appears spontaneously the vector $SU(2)$ symmetry is broken to $U(1)$ and two charged $\pi'$ mesons are expected to possess zero masses. In this case the matrix of second variation $ \hat V^{(2)}$ reads \ba \frac12 V^{(2)\sigma}_{11} &=& 4\lambda_1 \sigma_1^2  + 2\lambda_5 \sigma_1 \sigma_2 + 2 \lambda_4 \sigma_2^2 - 2 {\cal N} \sigma_1^2\ln{\frac{\mu+\sqrt{\mu^2-\sigma_1^2}}{\sigma_1}}, \no
 V^{(2)\sigma}_{12}
&=& 2 \lambda_5 \sigma_1^2  + 4 \lambda_3 \sigma_1 \sigma_2 + 2\lambda_6  \sigma_2^2, \qquad \frac12 V^{(2)\sigma}_{22} = 2 \lambda_4  \sigma_1^2 + 2 \lambda_6  \sigma_1 \sigma_2 + 4\lambda_2 \sigma_2^2 ,\no
 V^{(2)\sigma\pi}_{10} &=& \Big( 4 (\lambda_3 - \lambda_4)\sigma_1
+ 2\lambda_6\sigma_2 \Big) \rho,\qquad
 V^{(2)\sigma\pi}_{20} = \Big( 2 \lambda_6 \sigma_1
+ 8\lambda_2\sigma_2\Big) \rho, \no \frac12 V^{(2)\pi}_{00} &=& 4 \lambda_2\rho^2; \qquad \frac12 V^{(2)\pi}_{\pm\mp} = 0 , \label{secvarpi} \ea where the r.h.s. are evaluated with the help of Eqs.\gl{creq1} and \gl{creq2}. We notice that convexity around this minimum implies that all diagonal elements are non-negative $ V^{(2)\sigma}_{jj} > 0$. This gives positive masses for two scalar and one pseudoscalar mesons, whereas the triplet of pions and charged doublet of $\pi'$ mesons remain massless. Of course,  the mass spectrum can be obtained quantitatively  after kinetic terms are normalized.

Once we have fixed the interaction to quark matter we are not free in the choice of the kinetic term for scalar fields. Namely one cannot rotate two fields and rescale the field $H_1$ without changes in the chemical potential driver \gl{potmu}. However the rescaling of the field $H_2$ is possible at the expense of an appropriate redefinitions of other coupling constants and this freedom can be used to fix one of the constants which appear in the kinetic term. Thus we take the general kinetic term symmetric under $SU(2)_L\times SU(2)_R$ global rotations to be \be {\cal L}_{kin} = \frac14 \sum\limits^2_{j,k =1} A_{jk} \tr{\partial_\mu H^\dagger_j
\partial^\mu H_k} .
\ee After selecting out the v.e.v. $\langle H_1\rangle = \langle\sigma_1\rangle \equiv \bar\sigma_1$ one can separate the bare Goldstone boson action with the chiral parameterization \gl{chipar}.   Let us explore the kinetic part quadratic in fields. We expand $U = 1 + i \hat\pi/F_0 +\cdots,\ \xi = 1 + i \hat\pi/2F_0 +\cdots$ and use the v.e.v.'s $\sigma_j \equiv \bar\sigma_j + \Sigma_j$ $\hat\pi = \tau_3 \rho + \hat\Pi$. Then the quadratic part looks as follows \ba {\cal L}^{(2)}_{kin} &=& \frac12 \sum\limits^2_{j,k =1} A_{jk}\Biggl[
\partial_\mu \Sigma_j \partial^\mu \Sigma_k +\frac{1}{F_0^2} \bar\sigma_j \bar\sigma_k \partial_\mu \pi^a \partial^\mu \pi^a\Biggr]
\no&&+\frac{1}{F_0}\sum\limits^2_{j =1} A_{j2}\Biggl[- \rho
\partial_\mu \Sigma_j \partial^\mu \pi^0 +
  \bar\sigma_j  \partial_\mu \pi^a \partial^\mu \Pi^a\Biggr]
+ \frac12 A_{22}\Biggl[ \frac{\rho^2}{F^2_0}\partial_\mu \pi^0 \partial^\mu \pi^0 + \partial_\mu  \Pi^a \partial^\mu \Pi^a\Biggr]  \la{kinet}  ,
\ea which shows the mixture between light and heavy pseudoscalar states and, in the SPB phase, also between scalar and pseudoscalar states.

Let us define \be F_0^2 =  \sum\limits^2_{j,k =1} A_{jk} \bar\sigma_j \bar\sigma_k,\quad \zeta \equiv \frac{1}{F_0} \sum\limits^2_{j =1} A_{j2} \bar\sigma_j. \label{norm1} \ee In the symmetric phase $\rho = 0$, \gl{kinet} can be diagonalized, \ba {\cal L}^{(2)}_{kin,\pi} = \frac12 \partial_\mu \tilde\pi^a \partial^\mu \tilde\pi^a + \frac12 (A_{22} - \zeta^2)\partial_\mu \Pi^a \partial^\mu \Pi^a , \qquad A_{22} - \zeta^2 = \frac{\bar\sigma_1^2 \mbox{\rm det} A}{F_0^2} > 0 , \ea wherefrom, taking into account the matrix of second variations of the effective potential \gl{effpot1}, one finds the masses of the heavy pion triplet \be m^2_{\Pi} = \frac{- \Delta_{22} +(\lambda_3 - \lambda_4) (\bar\sigma_1)^2 + \lambda_6 \bar\sigma_1 \bar\sigma_2 + 2\lambda_2 (\bar\sigma_2)^2}{A_{22} - \zeta^2 }. \label{norm2} \ee

In the SPB phase the situation is more involved: pseudoscalar states mix with scalar ones. In particular, diagonalization is different for neutral and charged pions because the vector isospin symmetry is broken: $SU(2)_V \rightarrow U(1)$. The SPB induces mixing of both massless and heavy neutral pions with scalars. The (partially) diagonalized kinetic term has the following form \ba &&{\cal L}^{(2)}_{kin} = \partial_\mu \tilde\pi^\pm \partial^\mu \tilde\pi^\mp + \frac12 \Big( 1 +\frac{A_{22} \rho^2}{F^2_0} \Big)\partial_\mu \tilde\pi^0 \partial^\mu \tilde\pi^0  +  (A_{22} - \zeta^2)\partial_\mu \Pi^\pm \partial^\mu \Pi^\mp \no &&+\frac12 (A_{22} - \frac{F_0^2}{F_0^2 + A_{22} \rho^2}\zeta^2)\partial_\mu \Pi^0 \partial^\mu \Pi^0 \no &&+ \frac12 \sum\limits^2_{j,k =1} \frac{A_{jk} F_0^2 + \rho^2 \mbox{\rm det} A \delta_{1j}\delta_{1k}}{F_0^2 + A_{22} \rho^2}\partial_\mu \Sigma_j \partial^\mu \Sigma_k -\frac{F_0 \rho}{F_0^2 + A_{22} \rho^2}\zeta\partial_\mu \Pi^0 \sum\limits^2_{j =1} A_{j2}\partial^\mu \Sigma_j . \ea We see that even in the massless pion sector the isospin breaking $SU(2)_V \rightarrow U(1)$ occurs: neutral pions become less stable with a larger decay constant. Another observation is that in the charged meson sector the relationship between massless $\pi$ and $\Pi$ remain the same as in the symmetric phase.

\section{\label{Z-symmetry}\large\bf $P$-violation in models
with discrete $Z_2 \times Z_2$ symmetry.}
 As a relevant example we now examine models with residual discrete chiral symmetry (after the breaking $SU(2)_L \times SU(2)_R \rightarrow SU(2)_V$) under independent reflections $H_1 \rightarrow - H_1$ and/or $H_2 \rightarrow - H_2$ . Then  $\lambda_5 =\lambda_6 = 0, A_{12} = 0, \Delta_{12}= 0$, but $\lambda_4 \not= 0$. One can always fix $A_1 = A_2$ redefining the other parameters.

Let us now see how the general relations in the previous section are realized in this model. The analysis of Eqs. \gl{efeqs}, \gl{ineq1}, \gl{efeqs1} and \gl{ineq3} as well as the positivity of the second variation matrix leads to conclusion that in the symmetric phase the only solution compatible with the very possibility of $P$-breaking is $\sigma_2 = 0$ . In the SPB phase for these models the constraint \gl{creq2} $ 2 \lambda_4 \sigma_1 \sigma_2 = \Delta_{12} = 0$ has also a unique solution $\sigma_2 = 0$.  Therefore $\sigma_2 = 0$ everywhere. As for $\sigma_1$ we get for $\mu=0$ \be \sigma_1^2 = \frac{\Delta_{11}}{2 \lambda_1}.\nonumber \ee

The condition for extremum  when $\mu\neq 0$ reads \ba \Delta_{11} - 2 \lambda_1 \sigma_1^2 - (\lambda_3 - \lambda_4)\rho^2 \la{efeqsIII11}= {\cal N}\Theta(\mu-\sigma_1)\left[\mu\sqrt{\mu^2-\sigma_1^2}- \sigma_1^2\ln{\frac{\mu+\sqrt{\mu^2-\sigma_1^2}}{\sigma_1}}\right]   \ea and the condition for this extremum to correspond to  a SPB phase is now \be (\lambda_3 - \lambda_4) \sigma_1^2 + 2\lambda_2 \rho^2  = \Delta_{22}. \la{criteq} \ee

The second variation matrix for  $\sigma_2 = 0$ in both phases reads \ba \frac12 V^{(2)\sigma}_{11} &=& 2\Delta_{11} - 2(\lambda_3 - \lambda_4) \rho^2  - 2 {\cal N} \mu \sqrt{\mu^2-\sigma_1^2} > 0, \stackrel{\mu = 0}{\longrightarrow} 4\lambda_1 \sigma_1^2 , \no
 V^{(2)\sigma}_{12} &=& 0,
\qquad \frac12 V^{(2)\sigma}_{22} = -  \Delta_{22}  +  (\lambda_3 + \lambda_4)  \sigma_1^2 + 2\lambda_2 \rho^2  \stackrel{\mbox{\small SPB}}{\longrightarrow} 2 \lambda_4  \sigma_1^2 > 0 ,\no
 V^{(2)\sigma\pi}_{10} &=& 4 (\lambda_3 - \lambda_4)\sigma_1  \rho ,\qquad
 V^{(2)\sigma\pi}_{20} = 0,
\no \frac12 V^{(2)\pi}_{00} &=&  - \Delta_{22} +(\lambda_3 - \lambda_4) \sigma_1^2  + 6 \lambda_2\rho^2 \stackrel{\mbox{\small SPB}}{\longrightarrow} 4\lambda_2\rho^2 > 0, \no \frac12 V^{(2)\pi}_{\pm\mp} &=& - \Delta_{22} +(\lambda_3 - \lambda_4) \sigma_1^2 + 2 \lambda_2\rho^2 \stackrel{\mbox{\small SPB}}{\longrightarrow}  0. \label{secvarpiIII3} \ea As this matrix is positive definite one derives the following requirements for the low-energy hadronic model to make sense \ba \lambda_1 > 0,\quad \lambda_2 > 0,\quad \lambda_4 > 0,\quad \Delta_{11} > 0,\quad (\lambda_3 \pm \lambda_4)  \Delta_{11}
> 2 \lambda_1 \Delta_{22}, \label{ineq2III}
\ea where the last inequality is obtained from the positivity of $V^{(2)\sigma}_{22}, V^{(2)\pi}_{00}$ at zero $\mu$ .

As it follows from  \gl{homeq} there is now (at most) one critical point (within our approximations) where \be \sigma_1^2 = \frac{\Delta_{22}}{(\lambda_3 - \lambda_4)}. \la{efeqsIIIcrit1} \ee For this solution to correspond to a real value for $\sigma_1$ and therefore be acceptable, the r.h.s. of  \gl{efeqsIIIcrit1} has to be positive. In addition we need
 Eq.\gl{efeqsIII11} (after substituting $\rho=0$) to have
a solution for $\mu$ in the range $\sigma_1 \leq \mu < \infty$, \ba \frac{(\lambda_3 - \lambda_4)  \Delta_{11} - 2 \lambda_1 \Delta_{22}}{(\lambda_3 - \lambda_4)} = {\cal N}\left[\mu\sqrt{\mu^2-\sigma_1^2 } -\sigma_1^2 \ln{\frac{\mu+\sqrt{\mu^2-\sigma_1^2}}{\sigma_1 }} \right]
 >0.
 \la{efeqsIIIcrit2}
\ea
 Then after comparing inequalities in \gl{ineq2III}, \gl{efeqsIIIcrit1} and \gl{efeqsIIIcrit2} one concludes that \be \lambda_3 > \lambda_4 > 0,\quad  \Delta_{22} > 0 .\label{ineq2III+} \ee Thus  the SPB phase should exist  for all positive constants $\lambda_j >0, \Delta_{jj}>0$ if $\lambda_3 > \lambda_4$. The critical value of chemical potential can to be calculated from Eq. \gl{efeqsIIIcrit2}.

Now let us estimate the typical scales of $P$-breaking from meson spectroscopy. Since we have defined $A_1 = A_2$ we can find from Eq.\gl{secvarpiIII3} the ratios of masses.
 Just to get a feeling of the possible scales involved, let us make a tentative choice
 $$A_{11} = \frac19 = A_{22},\quad F_0 = 100 {\rm MeV},\quad \sigma_1
= 300 {\rm MeV} = 3 F_0,$$ according to the relation $F_0^2 = A_1 \sigma_1^2 $ , and use the units $F_0$ further on. As well \ba  m_{\pi} = 0,\quad m_{\Sigma_1} = 0.7 {\rm GeV} = 7 F_0 ,   \quad m_{\Pi} = 1.3 {\rm GeV}= 13 F_0,\quad m_{\Sigma_2} = 1.5 {\rm GeV} = 15 F_0 \ea in a fair agreement with particle phenomenology \cite{pdg}. Then from the definitions of masses \be m^2_{\Sigma_1} = \frac{2\Delta_{11}}{A_{11}} = \frac{4 \lambda_1 \sigma_1^2}{A_{11}};\quad m^2_{\Sigma_2} - m^2_{\Pi} = \frac{2\lambda_4 \sigma_1^2}{A_{11}} , \ee one finds
$$\Delta_{11} \simeq 2.7 F_0^2,\quad \lambda_1
\simeq 0.15,\quad \lambda_4 \simeq 0.35 .$$

Taking $\sigma_{1,crit} \simeq 1.8 F_0$ and the previously estimated value $\lambda_1 = 0.15$ one finds that SPB occurs at $p_F \simeq 3.9 F_0 = 1.44 p_{F,nuclear}$ which corresponds to dense nuclear matter with $\varrho_{B,crit}  \simeq 0.5 {\rm fm}^{-3} \simeq 3 \varrho_{B,nuclear} $. The phase transition occurs at $\mu_c \simeq 4.3 F_0 > \sigma_1$. From the definition of \gl{efeqsIIIcrit1} and of the mass of $\Pi$ one finds $\lambda_3 \simeq 3.6$ and $\Delta_{22} \simeq 11 F_0^2$ . Thus we see that the possibility of SPB emerges naturally for reasonable
 values of the meson physics parameters and low-energy constants. At this critical point the masses of scalar mesons are $m_{\Sigma_1}\simeq 1.7 F_0,\ m_{\Sigma_2}\simeq 4.5 F_0$ .

\section{Conclusions}
Let us summarize here our main findings. Parity breaking seems to be quite a realistic possibility in nuclear matter at moderate densities. We have arrived at this conclusion by using an effective lagrangian for low-energy QCD that retains the two lowest lying states in the scalar and pseudoscalar sectors. We include a chemical potential for the quarks that corresponds to a finite density of baryons and investigate the pattern of symmetry breaking in its presence. We have found the necessary and sufficient conditions for a phase where parity is  spontaneously broken to exist. In general this phase is bound and it extends across a
 range of chemical potentials that correspond to nuclear densities where
more exotic phenomena such as color-flavor locking or color superconductivity may occur.

Salient characteristics of this phase would be the spontaneous breaking of the vector isospin symmetry $SU(2)_V$ down to $U(1)$ and the generation two additional massless charged pseudoscalar mesons. We also find a strong mixing between scalar and pseudoscalar states that translate spontaneous parity breaking into meson decays. The mass eigenstates will decay both in odd and even number of pions simultaneously. Isospin breaking can also be visible in decay constants.

We think that our conclusions are drawn in a region of parameters where our effective lagrangian is applicable and, while obviously we cannot claim high accuracy in our predictions,  we are confident that the existence of this novel phase is not an spurious consequence of our approach but a rather robust prediction. It would surely be interesting to investigate how this new phenomenon could possibly influence the equation of state of neutron stars (the density of such objects seems to be about right for it).

Lattice methods could shed some light on this issue and confirm or falsify the existence of this interesting phase in dense nuclear matter.  One could probably use the expansions for small values of $\mu$ at finite temperatures to check some of our expressions. For this matching the natural approach is a hot hadron gas \cite{resongas} . Conversely, it would be possible to extend our techniques to the case of isospin chemical potential \cite{son}.

We are grateful to our collaborators S. Afonin and V. Andrianov for checking the calculations and making very useful remarks . This work was supported by research
grants FPA2007-66665, 2005SGR00564, 2007PIV10046. It is also supported by the
Consolider-Ingenio 2010 Program CPAN (CSD2007-
00042). We acknowledge the partial support of the EU
RTN networks FLAVIANET and ENRAGE and the Program
RNP2.1.1.1112.

\end{document}